\documentclass[conference]{IEEEtran}
\IEEEoverridecommandlockouts
\usepackage{cite}
\usepackage{amsmath,amssymb,amsfonts}
\usepackage{algorithm}
\usepackage{algorithmic}
\usepackage{comment}

\usepackage{graphicx}
\usepackage{subcaption}
\usepackage{textcomp}
\usepackage{hyperref}
\usepackage{makecell}
\usepackage{xcolor}
\def\BibTeX{{\rm B\kern-.05em{\sc i\kern-.025em b}\kern-.08em
    T\kern-.1667em\lower.7ex\hbox{E}\kern-.125emX}}
\begin{document}

\title{Charting the Uncharted: The Landscape of Monero Peer-to-Peer Network}

\author{\IEEEauthorblockN{1\textsuperscript{st} Yu Gao*}
\IEEEauthorblockA{ 
\textit{UZH Blockchain Center}\\
\textit{University of Zurich}\\
yugao@ifi.uzh.ch}
\and
\IEEEauthorblockN{2\textsuperscript{nd} Matija Pi\v{s}korec}
\IEEEauthorblockA{
\textit{University of Zurich}\\
\textit{Rudjer Boskovic Institute}\\
matija.piskorec@irb.hr}
\and
\IEEEauthorblockN{3\textsuperscript{rd} Yu Zhang}
\IEEEauthorblockA{
\textit{UZH Blockchain Center}\\
\textit{University of Zurich}\\
zhangyu@ifi.uzh.ch
}
\and
\IEEEauthorblockN{4\textsuperscript{th} Nicolò Vallarano}
\IEEEauthorblockA{
\textit{UZH Blockchain Center}\\
\textit{University of Zurich}\\
nicolo.vallarano@uzh.ch}
\and
\IEEEauthorblockN{5\textsuperscript{th} Claudio J. Tessone}
\IEEEauthorblockA{ 
\textit{UZH Blockchain Center}\\
\textit{University of Zurich}\\
claudio.tessone@uzh.ch}
}

\maketitle

\begin{abstract}

The Monero blockchain enables anonymous transactions through advanced cryptography in its peer-to-peer network, which underpins decentralization, security, and trustless interactions. However, privacy measures obscure peer connections, complicating network analysis. This study proposes a method to infer peer connections in Monero’s latest protocol version, where timestamp data is unavailable. We collect peerlist data from TCP flows, validate our inference algorithm, and map the network structure. Our results show high accuracy, improving with longer observation periods. This work is the first to reveal connectivity patterns in Monero’s updated protocol, providing visualizations and insights into its topology. Our findings enhance the understanding of Monero's P2P network, including the role of supernodes, and highlight potential protocol and security improvements.

\end{abstract}


\section{Introduction}
The peer-to-peer (P2P) network is the foundational layer of decentralized ledger technology (DLT) systems, ensuring decentralization, consensus, and network security through broadcast of transactions and blocks~\cite{eason1955certain,wood2014ethereum}. Although different DLT systems vary in design, all rely on the P2P network as the backbone for decentralization, security, and trustless interactions~\cite{bohme2015bitcoin,antonopoulos2023mastering,el2018review}. The P2P network consists of independent nodes (peers) that connect through an overlay structure to share resources. Each peer plays an equal role in sending messages and sharing resources~\cite{buford2009p2p}. The introduction of consensus and synchronization mechanisms adds complexity, affecting the network’s security, consensus, and smart contract vulnerabilities.


A key feature of P2P networks in DLT is dynamic discovery and interaction between nodes, forming a decentralized network structure governed by the P2P protocol~\cite{nakamoto2008bitcoin,marcus2018low}. This eliminates the need for central authority management, making the network resilient and scalable, driven by autonomous nodes. However, this self-organizing capability complicates the study of the network, requiring experimental measurements and tools such as "traceroute" to map the structure of the system\cite{eskandari2018first,tekiner2021sok}.

Monero, an open-source blockchain, uses advanced cryptographic techniques such as ring signatures, stealth addresses, and ring confidential transactions to ensure transaction anonymity on its permissionless P2P network. Monero employs Proof of Work (PoW) as its consensus mechanism but distinguishes itself by ensuring untraceable transactions. Although DLT research spans smart contracts~\cite{zou2019smart, zheng2020overview}, consensus mechanisms~\cite{bach2018comparative}, and decentralized finance~\cite{werner2022sok}, research on Monero's P2P network remains scarce, with studies focusing more on transaction traceability~\cite{moser2017empirical, li2019traceable}. However, the decentralized nature of Monero’s network has also made it a target for cryptojacking\cite{tekiner2021sok,eskandari2018first}, an illicit practice in which attackers exploit the computing power of users to mine Monero without consent. This issue not only challenges user privacy, but also poses security risks to the integrity of the network, highlighting the need for stronger safeguards in Monero's P2P ecosystem.

Few studies investigate Monero’s P2P network, the most notable being Cao et al.~\cite{cao2020exploring}, which notes an update of the protocol to hide the \textit{time\_stamp} field in peer lists. This update removes the timestamp information, making it harder to infer peer connections. Existing studies on Monero focus mainly on traceability~\cite{moser2017empirical}, highlighting the need for more research on its P2P network.

In this paper, we analyze Monero’s P2P network by collecting peer lists extracted from the TCP data exchanged between our Monero node and its peers. To overcome the absence of timestamps, we propose an algorithm that infers peer connections using timestamp-free TCP packet analysis. We validate this approach by comparing the inferred connections with the actual neighbors, achieving high accuracy. Our analysis of Monero's network topology provides insights into the structure of the system, potential vulnerabilities, and areas for future improvements in protocol design.

\section{Monero Peer-to-Peer Protocol}

By default, a Monero node establishes 8 outgoing connections. At least one outgoing connection is required to join the network (no such requirement for incoming connections). When a new node joins, it connects to the public seed nodes to discover other active nodes. Through a handshake protocol, the node establishes connections, validates them, and synchronizes data such as blocks and peer lists, while maintaining and updating its peer list, managing connection requests, and ensuring network stability.

Unlike Bitcoin-based P2P protocols, Monero uses its unique protocol. Each peer has two lists: \textit{white\_list} (1000 slots) for recent handshakes and \textit{grey\_list} (5000 slots) for unresponsive peers or those with lower timestamp rankings. Under the new Monero protocol, a peer returns 120\% of peers from its top 300 \textit{last\_seen} peers in the white list during the TCP flow process\footnote{\href{https://github.com/monero-project/monero/blob/960c2158010d30a375207310a36a7a942b9285d2/src/p2p/net_peerlist.h}{Monero GitHub Repository}}.

\section{Peer list collection and neighbor inference}

In P2P networks within DLT systems, it is important to differentiate between ``active neighbors'' and ``potential peers''.

\subsection{Definition of ``Neighbor'' and ``Interaction''}

To formalize this distinction, we define a ``connection'' and an ``interaction'' as follows:

\textbf{Neighbor/Connection:} A connection is a persistent, active communication channel between peers, allowing the exchange of data. Active neighbors are peers with ongoing, direct connections, essential for maintaining network integrity and performance as peers join and leave over time.

\textbf{Interaction:} An interaction refers to a temporary relationship between peers, including:
(1) Temporary connections, such as during initial handshakes, without forming a persistent channel.
(2) Potential peers listed in a node peer list, based on discovery protocols, without established communication or validation.
This distinction helps to accurately represent and manage the network topology, ensuring that only meaningful interactions are considered active neighbors.

To infer P2P connections between Monero nodes, we first set up a peer list data collection pipeline, as outlined in~\cite{cao2020exploring}, and then develop a new inference algorithm. Due to changes in Monero's peer list sharing protocol, which no longer includes the \textit{last\_seen} timestamp, we propose a modified algorithm that works without these data.

\subsection{Peer list data collection}
\begin{figure}[htbp]
\centerline{\includegraphics[width=0.4\textwidth]{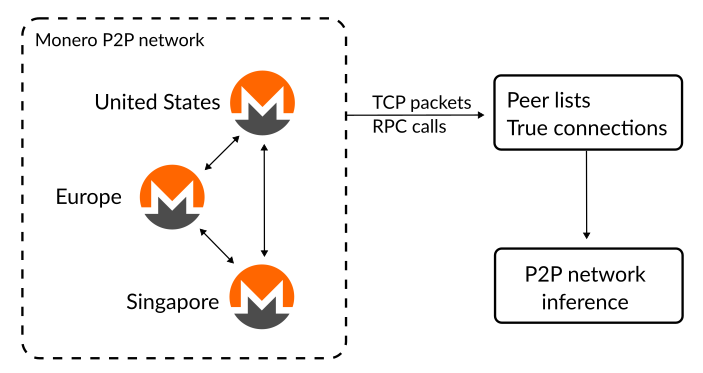}}
\caption{Data collection pipeline.}
\label{fig:data-collection}
\end{figure}

Our data collection pipeline (Figure~\ref{fig:data-collection}) consists of three machines deployed in the United States, Europe, and Singapore. Geographical dispersion is crucial to capturing a broad view of the Monero P2P network, while the controlled nodes allow for validation of the P2P inference. Peer list data is collected by running a Monero node and monitoring TCP traffic with the \textit{tcpflow}\footnote{\href{https://github.com/simsong/tcpflow}{https://github.com/simsong/tcpflow}} program, listening on port 18080, through which communication with other nodes occurs. Peer lists are extracted from the binary dump of the traffic data, as they are not exposed through the node's RPC interface.

The connection data from our three nodes is used to validate the P2P network inference. Data for this paper were collected over three weeks, from 21.12.2024 to 10.1.2025.

\subsubsection{Identifying Real Neighbors}

To identify real neighbors, we analyze peer list data based on the P2P protocol, where each peer shares up to 250 entries from its top 300 whitelist peers. Real neighbors, which maintain frequent handshakes, appear more often in shared \textit{peer\_lists} than non-neighbors. The nodes belong to one of two categories:
\begin{itemize}
    \item \textbf{Highly connected nodes}: for example, public seed nodes with numerous connections.
    \item \textbf{Typical nodes}: default 8 outgoing neighbors.
\end{itemize}

The challenge is to set a frequency threshold to distinguish real neighbors. Since peer selection is bidirectional and is governed by the \texttt{IDLE\_HANDSHAKE} protocol, real neighbors are expected to appear consistently in updated whitelists.

For a node \( i \), we define the relative presence of an observed address \( a \) as:
\begin{equation*}
    p_i(a) = \frac{\sum_{j=1}^{n_i} \chi(a \in P_j)}{n_i}
\end{equation*}
where \( P^{(i)} \) represents \( i \)'s received TCP packets and \( \chi \) is an indicator function.

A node with 8 neighbors, receiving 250 peer addresses per handshake, selects a neighbor with probability \( P_{\text{neighbour}} \approx 0.833 \). For non-neighbors, selection follows two stages:
\begin{enumerate}
    \item Entering the top 300 with probability \( P_{\text{enter}} \approx 0.302 \).
    \item Being chosen within the top 300 with \( P_{\text{selected}} \approx 0.833 \).
\end{enumerate}
Thus, the overall probability for random nodes is:
    $P_{\text{random}} = P_{\text{enter}} \cdot P_{\text{selected}} \approx 0.252$.

This structural bias results in real neighbors appearing three times more frequently than random nodes, making them distinguishable through cumulative frequency analysis.

\subsection{Neighbor identification by $k$-means}
The simple method above intuitively identifies high-frequency peer interactions compared to low-frequency ones. This observation led us to apply $k$-means clustering to separate high- and low-frequency links.

To identify real neighbors, we first collect the frequency of each peer's connections in shared \textit{peer\_lists} or TCP flows, then use the $k$-means clustering algorithm to differentiate between high-frequency and low-frequency links.

\begin{algorithm}
\caption{Data Filtering and Neighbor Inference Process}
\label{alg:filtering}
\begin{algorithmic}[1]
\small
\REQUIRE Raw dataset $D$ with columns \texttt{ip1}, \texttt{ip2}, \texttt{count}.

\STATE $C_{\text{min}} \gets 2$, threshold of minimum connection count.
\STATE $N_{\text{min}} \gets 8$, threshold of minimum number of out-going neighbors.

\STATE $D_{\text{filtered}}=\{row \in D: row\texttt{[count]}\geq C_{\text{min}}\}$

\FOR{each unique source IP $s$ in $D_{\text{filtered}}$}
    \STATE $G_s=\{row \in D_{\text{filtered}}: row\texttt{[ip1]}=s\}$
    \IF{$|G_s| \geq N_{\text{min}}$}
        \STATE Perform $k$-means clustering on the unique $\texttt{count}$ values in $G_s$ with $k = 2$. Greater $\texttt{count}$ values' \texttt{label}s are set as 1 and as 0 otherwise.
    \ENDIF
\ENDFOR
\STATE $D'=\{row \in D_{\text{filtered}}: row\texttt{[label]}=1\}$
\STATE \textbf{return} $D'$
\end{algorithmic}
\end{algorithm}

The dataset consists of triplets \((\texttt{ip1}, \texttt{ip2}, \texttt{count})\), where \(\texttt{ip1}\) and \(\texttt{ip2}\) are IP addresses, and \(\texttt{count}\) is the interaction frequency between them. We excluded rows where \(\texttt{ip1}\) corresponds to Singapore, the US, or the EU, since our nodes perform frequent handshakes. To identify relevant connections and infer real neighbors, we applied the following steps:

First, we remove the rows where \(\texttt{count} \leq 1\), as these likely represent noise, leaving the dataset \(D_{\text{filtered}}\) with meaningful interactions.

Second, we grouped the data by \(\texttt{ip1}\), creating groups \(G_s\) for each source IP. For each group:\\
- We discarded groups with fewer than two interactions.\\
- We applied $k$-means clustering with \(k = 2\) on the unique \(\texttt{count}\) values, which distinguishes true connections from non-true connections based on frequency.

Finally, we retained all connections labeled as true neighbors (for example, those clustered as 1), forming the dataset \(D'\), which includes the most relevant inferred connections.

By clustering the frequency data, we identified high-frequency peers as likely true neighbors, with the $k$-means method offering computational efficiency and revealing meaningful patterns in large-scale interaction datasets.

\subsection{Neighbor inference validation}
We compare the inferred neighbors of our Singapore, EU, and US nodes with the real connection lists obtained using Monero's \href{https://www.getmonero.org/resources/developer-guides/daemon-rpc.html#get_connections}{monero-daemon-rpc}. These real connection lists serve as a benchmark for validating the neighbor inference.



Classifier quality measures like precision and recall help assess our method's performance. Precision is the ratio of true positives to all detected positives (true + false positives), representing the proportion of real neighbors correctly identified. Recall is the ratio of true positives to all actual neighbors, indicating the proportion of real neighbors successfully detected.

Since peers dynamically connect and disconnect over time, even real connection lists capture only a transient snapshot of the network. To validate our approach, we first identify neighbors of our node in the inferred network and then check how many of these inferred neighbors appear in the real connection lists during data collection.

We measure precision as the ratio of ``our node’s neighbors in the inferred network'' to ``neighbors identified from real connections over one week of data.'' The following tables present the validation results. 

\begin{table}[htbp]
\caption{One-week data neighbor inference accuracy}
\begin{center}
\begin{tabular}{|c|c|c|c|}
\hline
\multicolumn{4}{|c|}{\textbf{\makecell{Benchmark comparison for one week data }}} \\
\hline
\textbf{\textit{Location}} & \textbf{\textit{Singapore}} & \textbf{\textit{US}} & \textbf{\textit{EU}} \\
\hline
\makecell{Inferred network neighbors} &  489 & 534 & 534\\
\hline
\makecell{Neighbors in connection list} & 388 & 367& 371\\
\hline
\makecell{Neighbor identification Precision} & 79.35\% & 68.73\%& 69.48\%\\
\hline
\end{tabular}
\label{tab1}
\end{center}
\end{table}

\begin{table}[htbp]
\caption{Three weeks data neighbor inference accuracy}
\begin{center}
\begin{tabular}{|c|c|c|c|}
\hline
\multicolumn{4}{|c|}{\textbf{\makecell{Benchmark comparison for three weeks data}}} \\
\hline
\textbf{\textit{Location}} & \textbf{\textit{Singapore}} & \textbf{\textit{US}} & \textbf{\textit{EU}} \\
\hline
\makecell{Inferred network neighbors} &  394 & 437 & 713\\
\hline
\makecell{Neighbors in connection list} & 327 & 319 &  572\\
\hline
\makecell{Neighbor identification Precision} & 82.99\% & 73.00\%& 80.22\%\\
\hline
\end{tabular}
\label{tab2}
\end{center}
\end{table}

Without incorporating the \textit{last\_seen} timestamps, the prediction accuracy of our inference method is already relatively high. This result in Table \ref{tab1} is based on only one week of listening data, excluding the \textit{last\_seen} timestamps in the peer lists under the current protocol. When the data collection period was extended, the accuracy of the EU node improved by approximately 10\% (check Table \ref{tab2}), highlighting the significant impact of a larger dataset collected over a longer period on the prediction accuracy. This improvement aligns with the theoretical expectation that longer TCP listening times enhance accuracy.

\section{Monero P2P Network Topology Analysis}

After applying $k$-means clustering to filter the links, we obtained the ``high-frequency clusters'' edge list and mapped it into a network.

\subsubsection{Network Topology Analysis}

P2P networks in DLT systems, defined by their decentralized architecture, present both challenges and opportunities that require a thorough analysis. Network Science, grounded in graph theory and statistical physics, offers a solid framework for studying the structural and functional properties of networks \cite{newman2018networks,watts1999networks}. By representing a complex system as a graph, we can analyze topological features, develop network models, and evaluate robustness.


Although visualizing the network provides an initial insight, this approach becomes limited for larger networks. Therefore, a qualitative analysis of the network features using statistical data is essential. We define centrality and betweenness centrality as follows \cite{newman2018networks}:

\textbf{Degree Centrality}: Measures the number of direct connections (edges) a node has. The degree \( k_i \) of the node \( i \) is defined as:
$k_i = \sum_{j=1}^{n} A_{ij}$,
where \( A_{ij} \) is the adjacency matrix of the network. Degree centrality is useful for analyzing network centralization and structure.

\textbf{Betweenness Centrality}: Quantifies how often a node lies on the shortest paths between pairs of nodes, indicating its influence on network interactions. The betweenness centrality \( b_i \) of node \( i \) is given by:
$b_i = \sum_{st} \frac{n_{st}^i}{g_{st}}$,
where \( n_{st}^i \) is the number of shortest paths from \( s \) to \( t \) passing through \( i \), and \( g_{st} \) is the total number of shortest paths between \( s \) and \( t \). 

\begin{figure}[htbp]
\centerline{\includegraphics[width=0.45\textwidth]{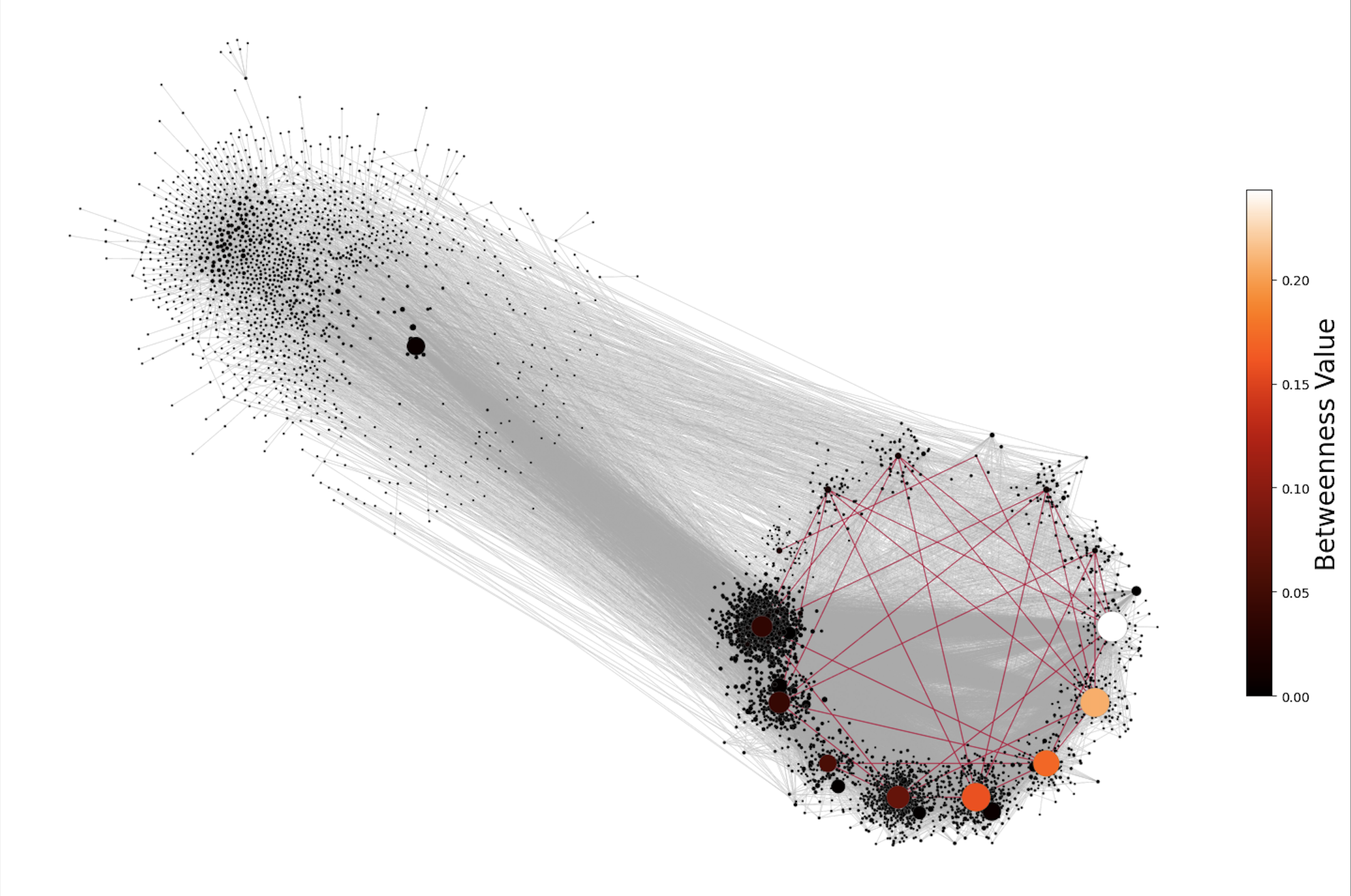}}
\caption{Visualization of the LCC, with node colors representing betweenness centrality and node sizes proportional to their degree centrality.}
\label{fig network}
\end{figure}

Fig. \ref{fig network} visualizes the LCC of the Monero P2P network, where node color represents betweenness centrality and size reflects degree centrality. Warmer-colored nodes indicate higher betweenness, highlighting their role in connecting different parts of the network, while larger nodes signify high-degree peers. The network consists of 4,837 nodes, with 3,153 directly linked to the 14 top-degree nodes, collectively covering 82.1\% of the network. These core nodes—primarily public seed nodes or supernodes—maintain multiple connections, facilitating efficient data exchange and ensuring network integrity.

\subsubsection{Interconnection Analysis Among Supernodes}

\begin{figure}[htbp]
    \centering
    \includegraphics[width=0.9\linewidth]{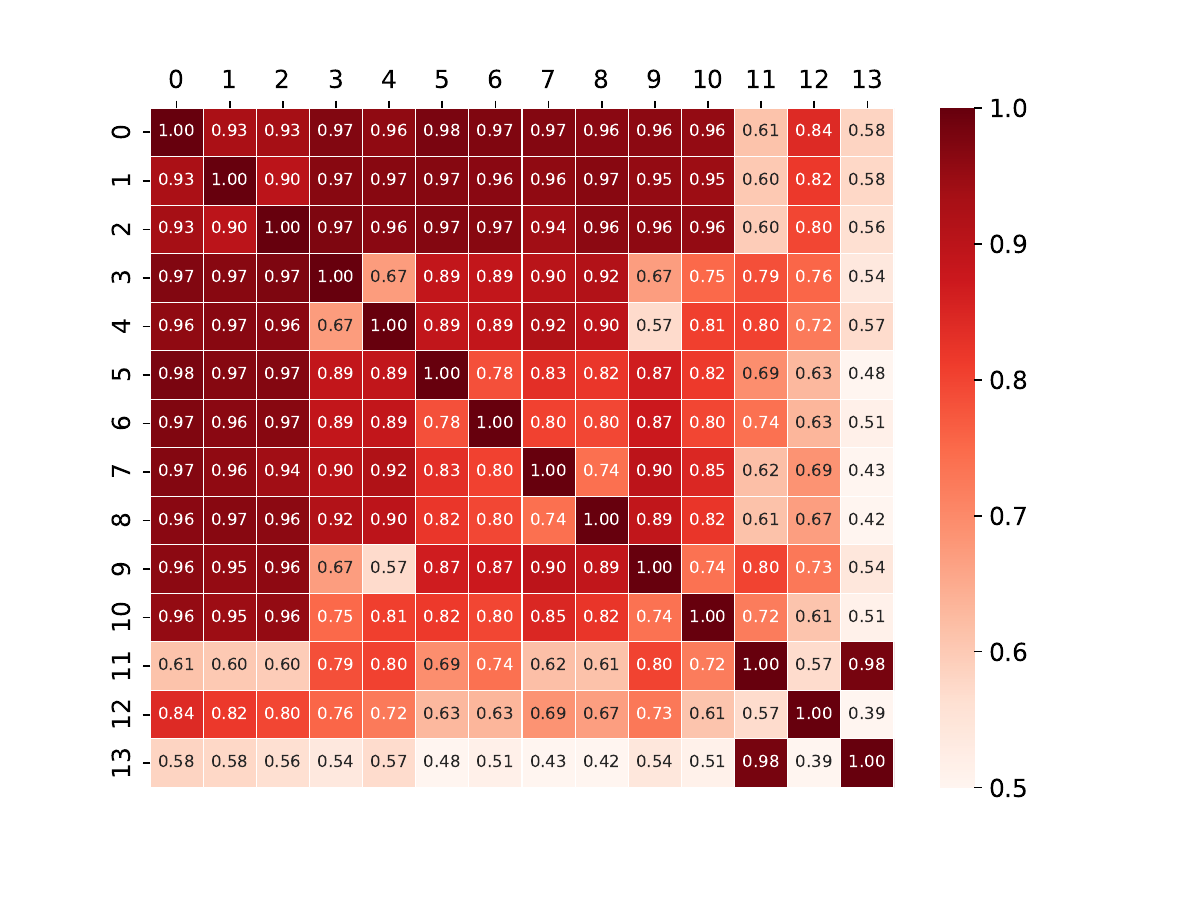}
    \caption{One-hop neighbor overlap rate among the 14 top-degree nodes. The value in each cell represents the overlap rate of the one-hop neighbors between the nodes indexed by the X and Y ticks.}
    \label{fig: heatmap}
\end{figure}

To explore the interconnections among these supernodes, we examine the overlap of their direct neighbors. The heatmap in Fig. \ref{fig: heatmap} shows the overlap rate of direct neighbors among the 14 top-degree nodes. The value in each cell represents the overlap rate of the one-hop neighbors between the nodes indexed by the X and Y ticks, calculated as:
$r = \frac{n_{overlap}}{\min(n_x, n_y)}$,
where \( n_x \) and \( n_y \) are the number of neighbors of nodes \( x \) and \( y \), and \( n_{overlap} \) is the number of shared neighbors. The heatmap reveals that over 91\% of each top-degree node's direct neighbors are connected to other top-degree nodes, with this proportion nearing 100\% for 9 out of the 14 nodes. This suggests a strong level of interconnectivity and resilience within the core of the network.

\subsubsection{Robustness of the Monero P2P Network}

The absence of a central node is a defining feature of DLT systems. However, this does not preclude the existence of supernodes within the network. In an unstructured DLT P2P network, if small-world characteristics emerge from user social relationships, nodes with significantly higher degrees can exert disproportionate influence on the network structure, introducing potential risks such as Eclipse attacks \cite{henningsen2019eclipsing, delmolino2016step}, Sybil attacks \cite{bissias2014sybil, asfia2019blockchain}, and DoS attacks \cite{garber2000denial}.

\begin{figure}[htbp]
\centerline{\includegraphics[width=0.5\textwidth]{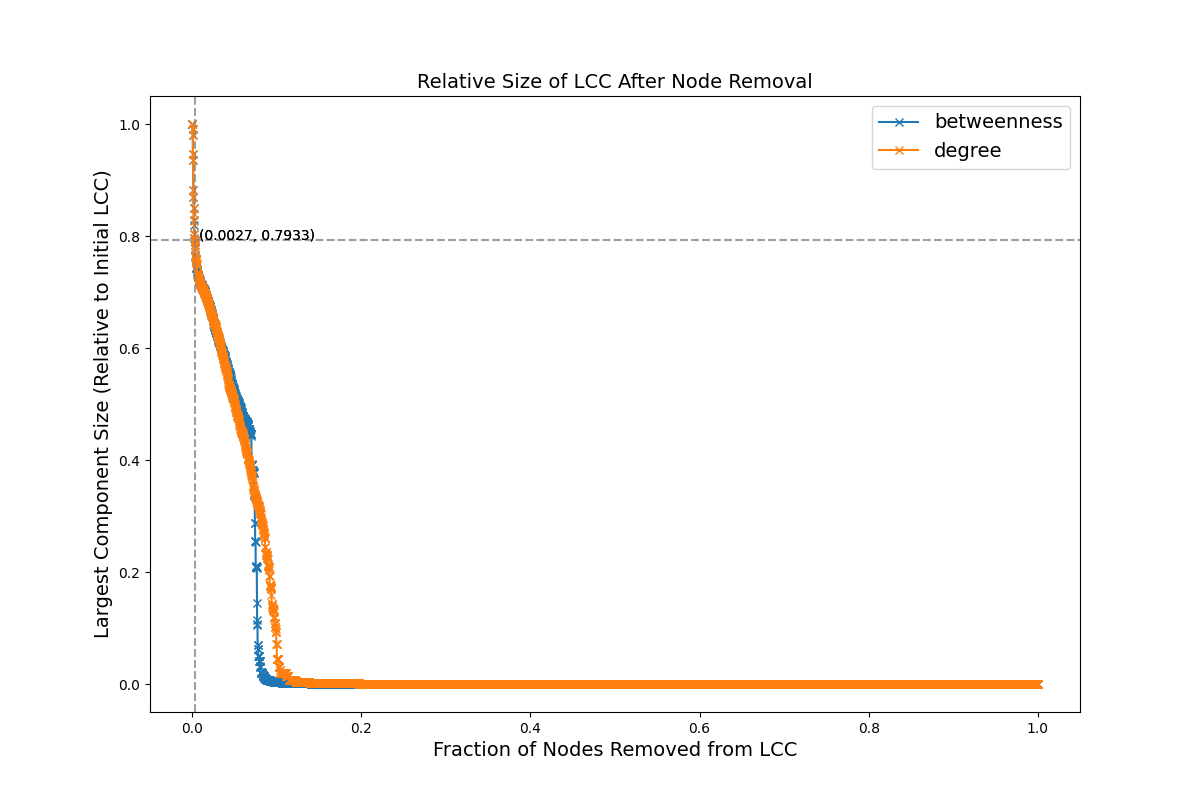}}
\caption{LCC attack by removing high betweenness and degree centrality nodes, where the turning point of LCC $\approx 0$ are 0.094 and 0.12 for betweenness centrality and degree centrality.}
\label{fig attack}
\end{figure}

To assess the network robustness, we focus on high-degree and high-betweenness centrality nodes and evaluate how quickly the largest connected component (LCC) disintegrates. As shown in Fig. \ref{fig attack}, the network demonstrates strong resilience. After removing the first 14 supernodes, the LCC size decreases by 20\%. In particular, the LCC collapses to nearly zero when around 9.4\% and 12\% of nodes are removed based on betweenness and degree centrality, respectively. This suggests that compromising these nodes or their neighbors would significantly disrupt the network, highlighting its inherent robustness.

\section{Conclusion}

In conclusion, this study introduces a novel method for mapping the Monero peer-to-peer network by identifying real neighbors between nodes. Our approach accurately uncovers the network's structure and provides an intuitive visualization of its topology. The findings offer the first comprehensive insights into connectivity patterns within the Monero P2P network under the new peer protocol.

Our results show that the network is highly centralized around several super nodes with significant betweenness centrality and high degrees. While this centralization strengthens security and robustness, it also introduces potential vulnerabilities. This study provides a foundation for future improvements to the Monero peer protocol, enhancing security and decentralization.


\bibliographystyle{IEEEtran}
\bibliography{reference}


\end{document}